\documentstyle[12pt,epsfig]{article}
\def\bmP{{\bf P}}
\def\bmV{{\bf V}}
\def\bmJ{{\bf J}}
\def\bmS{{\bf S}}
\def\bmq{{\bf q}}

\def\H3{$^3$H}
\def\4He{$^4$He}
\def\He3{$^3$He}
\def\haf{\textstyle{1\over2}}
\def\minus{\mbox{$-$}}
\def\bmmu{\mbox{\boldmath$\mu$}}
\def\bms{\mbox{\boldmath$\sigma$}}
\def\simeq{\mathrel{\raise4pt\hbox to 9pt{\raise -5pt\hbox{$=$}\hss{$\sim$}}}}

\begin{document}
\vspace*{-1.0in}
\hspace{\fill} \fbox{\bf LA-UR-97-2330}

\begin{center}

{\Large {\bf Nuclear Sizes and the Isotope Shift}}\\

\vspace*{0.80in}

J.L.\ Friar \\
Theoretical Division \\
Los Alamos National Laboratory \\
Los Alamos, NM  87545 USA\\

\vspace*{0.15in}

and

\vspace*{0.15in}

J.\ Martorell\\
Departament d'Estructura i Constituents de la Materia\\
Facultat F\'isica\\
Universitat de Barcelona\\
Barcelona 08028 Spain\\

\vspace*{0.15in}

and

\vspace*{0.15in}

D.\ W.\ L.\ Sprung\\
Department of Physics and Astronomy\\
McMaster University\\
Hamilton, Ontario, L8S 4M1 Canada\\

\end{center}

\vspace*{0.50in}

\begin{abstract}
Darwin-Foldy nuclear-size corrections in electronic atoms and nuclear radii
are discussed from the nuclear-physics perspective. Interpretation of precise
isotope-shift measurements is formalism dependent, and care must be exercised
in interpreting these results and those obtained from relativistic electron 
scattering from nuclei. We strongly advocate that the entire nuclear-charge 
operator be used in calculating nuclear-size corrections in atoms, rather than
relegating portions of it to the non-radiative recoil corrections. A preliminary
examination of the intrinsic deuteron radius obtained from isotope-shift 
measurements suggests the presence of small meson-exchange currents (exotic
binding contributions of relativistic order) in the nuclear charge operator, 
which contribute approximately $\haf$\%.
\end{abstract}
\pagebreak

Recent measurements in Garching \cite{1} and Paris \cite{2} have  
greatly improved our knowledge of the isotope shift between deuterium and 
normal hydrogen.  Due to their much increased precision\cite{3},
these measurements now rival the traditional relativistic electron  
scattering \cite{4} for determining the (nuclear) sizes of these isotopes  
(and their differences).  This new level of precision has led to a  
reexamination of many contributions to the level shifts \cite{5,6} and to the
calculation of higher-order QED processes.  Inevitably, a certain  
amount of controversy has ensued over the best way to proceed and over  
the proper interpretation of various mechanisms\cite{5,6}. Our  
purpose here is to discuss these topics briefly from the nuclear-physics  
perspective, given that these measurements have presented nuclear  
physics with great opportunities.  Nothing that we say here is entirely  
new (indeed, much is very old \cite{4,7,8}), but we believe that the  
totality casts considerable light on the interpretation and significance of  
these measurements.

Specifically, (1) we will (briefly) review the physics from the nuclear-%
physics perspective. (2) We will discuss the conventions (formalism  
dependence) attendant to introducing nuclear size.  Although there is  
no right or wrong way to do this, there are consistent or inconsistent  
ways to proceed and there are ample opportunities for double counting.  
(3) We will make recommendations for avoiding such problems and discuss
recent electron-scattering results\cite{9,10,11} from this perspective.
(4) We will make a first assessment of the d-p isotope-shift data in terms 
of ``normal'' and ``exotic'' components of  
the deuteron structure, even though the latter are not yet entirely well  
defined \cite{12}.  A new generation (``second generation'') of nuclear  
potentials \cite{13,14,15} gives improved insight into deuteron  
structure, and this will prove useful in reducing theoretical uncertainties.

Relativistic electron scattering has traditionally been the only  
successful method for measuring the sizes of the lightest nuclei  
\cite{4}.  Muonic atoms provided significant information on heavier  
nuclei but until very recently electronic-atom measurements lacked  
the necessary precision.  Nuclear physics has been investigated  
primarily using nonrelativistic dynamics, but the increasing precision  
of electron-scattering data in the late 1960's and early 1970's led to a  
reexamination \cite{7,8,12} of the ways that relativity can affect a  
nuclear charge distribution.  In order to be as specific as possible, we  
will first discuss various options that have arisen in discussing the simpler  
and better-known proton charge distribution, and then extend the  
discussion to light nuclei.  We use natural units ($\hbar = c = 1$) and the 
conventions and metric ($p^2 = m^2$) of Ref.\ \cite{16}. We also remove the
proton charge, $e_p$, from all currents.

For historical reasons (analogy with the electron) the electromagnetic  
structure of the proton was introduced in terms of two form factors  
(i.e., Lorentz scalars): the Dirac form factor, $F_1 (q^2)$, and the Pauli  
(anomalous magnetic moment) form factor, $F_2 (q^2)$.  The  
covariant current (normalized to unit charge) is given by \cite{16}
$$
J^{\lambda} = \bar{u} (\bmP^{\prime})(\gamma^{\lambda} F_1 (q^2) +  
\frac{i \kappa_p}{2M} F_2 (q^2)\, \sigma^{\lambda \nu} q_{\nu}) u (\bmP)\, ,
\eqno (1)
$$
where $\gamma^{\lambda}$ and $\sigma ^{\lambda \nu}$ are Dirac matrices, 
$u(\bmP)$ and $u(\bmP^\prime)$ are Dirac spinors, $\kappa_p$ is the proton 
anomalous 
magnetic moment, $M$ is the nucleon mass, $F_1 (0) = F_2 (0) = 1$, and $q =  
(P^{\prime} - P)$ is the momentum transferred (by an electron) to the  
final nucleon $(P^{\prime})$ from the initial one $(P)$.  Because $q^2 < 0$ 
for scattering kinematics, it is convenient to adopt the SLAC  
convention, $Q^2 \equiv -q^2 > 0$, thus avoiding inconvenient minus signs.

It was soon realized that even though $F_2$ primarily describes magnetic 
properties of the nucleon, it also contributes (in a minor way at small $Q^2$) 
to the charge distribution\cite{17}, so the Sachs\cite{18} charge 
and magnetic form factors, $G_E$ and $G_M$, respectively, were introduced:
\begin{eqnarray*}
\hspace*{1.5in} G_E &=& F_1 (Q^2) - \frac{\kappa_p Q^{2}}{4 M^{2}}  
F_2 (Q^2)\, , \hspace*{1.90in} (2a) \\
G_M &=& F_1 (Q^2) + \kappa_p F_2 (Q^2)\, .  
\hspace*{2.17in} (2b)
\end{eqnarray*}
In terms of these form factors, the (laboratory-frame) cross section for  
(massless) electron scattering by protons in first-Born approximation is  
given by the Rosenbluth formula \cite{19,4,17}
$$
\frac{d \sigma}{d \Omega} = \sigma_{\rm Mott} \left[ A_0 (Q^2) + B_0  
(Q^2) \left( \frac{1}{2} + \left( 1 + \frac{Q^{2}}{4 M^{2}} \right) \tan  
^2(\theta/2) \right) \right]\, , \eqno (3)
$$
where $\theta$ is the electron scattering angle, $\sigma_{\rm Mott}$ is the  
cross section for a spinless point particle, and
\begin{eqnarray*}
\hspace*{1.5in} A_0 (Q^2) &=& \frac{G^{2}_{E} (Q^{2})}{1 +  
\frac{Q^{2}}{4 M^{2}}} \equiv \widetilde{G}^{2}_E\, , \hspace*{2.15in}  
(4a) \\
B_0 (Q^2) &=& \frac{Q^2}{2 M^2} \left[\frac{G^2_M (Q^2)}
{1 + \frac{Q^2}{4 M^2}} \right] \equiv \frac{Q^2}{2  
M^2} \widetilde{G}^2_M \, . \hspace*{1.20in} (4b)
\end{eqnarray*}
Equation (3) applies to elastic electron scattering by an arbitrary  
nucleus, while Eq.\ (4) applies only to spin-$\haf$ systems (such as the  
proton, \He3, or \H3 ).  The form factors, $\widetilde{G}_E$ and  
$\widetilde{G}_M$ were proposed long ago\cite{17,20,4} as alternatives to 
$G_E$ and $G_M$, but were never popularly adopted.  Equation (3)  
has been written so that $A_0$ is a form factor associated with the  
charge distribution, while $B_0$ is analogously associated with the  
magnetization distribution obtained from the transverse (to  
${\bf \hat{q}}$) component of the (space) current.  This  
division is most transparently performed in Coulomb gauge \cite{7}.   
Often the square bracket in Eq.\ (3) is rearranged as [$A(Q^2)
+ B(Q^2) \tan^2 (\theta / 2)$], but then $A$ is no longer  
associated solely with the proton charge distribution.

One has the option of describing the proton's structure in terms of  
$(F_1, F_2)$, $(G_E, G_M)$, or $(\widetilde{G}_E, \widetilde{G}_M)$.  
Only the last  
option correctly gauges the proton charge distribution to order $(v / c)^2$ 
(or, equivalently, $Q^2 / M^2$).  Factors of $\tau = Q^2 / 4M^2$ and $\eta = 1 
+ \tau$ are of relativistic origin and also affect the proton mean-square  
charge radius, defined in the Breit frame\cite{7,17} as $\langle r^2 
\rangle_{\rm ch} \equiv \int d^3 x\, x^2 \, \rho(x)$, where $J^{\lambda}=(\rho,
{\bf J})$. Further defining 
$\langle r^2 \rangle_1 = -6\, F^{\prime}_{1} (0)$ and
$\langle r^2 \rangle_E = -6\, G^{\prime}_{E} (0)$, we obtain from Eq.\ (2a)
$$
\langle r^2\rangle_E = \langle r^2 \rangle_1 +  
\frac{3 \kappa_p}{2 M^{2}}\, , \eqno (5a)
$$
while the charge form factor obtained from Eq.\ (4a) produces
$$
\langle r^2\rangle_{\rm ch} = \langle r^2 \rangle_E +  
\langle r^2 \rangle_{\rm DF}\, , \eqno (5b)
$$
where we have defined $\langle r^2 \rangle_{\rm ch} = -6\, 
\widetilde{G}^{\prime}_{E} (0)$ and
$$
\langle r^2\rangle_{\rm DF} = \frac{3}{4 M^{2}}\, . \eqno (5c)
$$
The various mean-square radii, $\langle r^2\rangle_1$, $\langle r^2\rangle_E$, 
and $\langle r^2\rangle_{\rm ch}$, differ by amounts of order  
$\left(\frac{1}{M^{2}} \right) \sim 0.044\, {\rm fm}^2$, but are formally  
identical in the nonrelativistic (large-$M$) limit. Note that $\langle 
r^2\rangle_E^{1/2}$ is often called the proton radius, $r_p$\cite{20x}.

The quantity ($3/4M^2$) in Eq.\ (5c) is the Darwin-Foldy (DF) 
term \cite{16,21} and is obtained by expanding the $1/\eta$ factor in  
Eq.\ (4a).  This factor is traditionally incorporated into the  
kinematical factors (along with $\sigma_{\rm Mott}$) and the experimental data 
are then used to determine $G_E$ and $G_M$.  That is,  
{\bf by convention}, the Darwin-Foldy term is not considered part of the 
proton structure, even though it affects the cross section.

Nevertheless, to order ($1/M^2$) we can easily expand the $\lambda = 0$  
component of Eq.\ (1) to obtain the true charge density.  One finds  
that the {\bf covariant} form of $u$ (normalized to  
$\bar{u} u = 1$) generates a frame-dependent total charge (obtained  
by setting $q \rightarrow 0$).  The reason for this is that the wave  
function normalization factor $\left( \frac{1}{\sqrt{2E}} \right)$  
appropriate for this convention is relegated to the phase space $\left(  
{\rm i.e.,} \frac{d^{3} P}{(2E)(2 \pi)^{3}} \right)$.  If on the other  
hand, we incorporate that factor in $J^{\lambda}$, the phase space is  
$\left( \frac{d^{3}P}{(2 \pi)^{3}} \right)$ and the total charge is  
{\bf invariant}\cite{7,8}.  The invariant form of the charge 
operator\cite{16,21} is
$$
\rho \simeq \left( 1 - \frac{\bmq^{2}}{8 M^{2}} \right) G_E + i  
\frac{(2 G_{M} - G_{E})}{4 M^{2}} \; \bms \; \cdot \; \bmq \; \times  
\; \bmP, \eqno (6) 
$$
where the Darwin-Foldy factor $(\bmq^{2}/8M)$ is an explicit part of the charge 
operator, as is the spin-orbit interaction (expressed here in terms of the 
Pauli spin operator, $\bms$). The spin-orbit interaction plays a significant 
role in the isotopic charge-density differences of heavier nuclei\cite{4,21x}. 
Equation (6) for the charge distribution is equivalent (to ${\cal O} (1/M^2)$) 
to using the form factor $\widetilde{G}_E$. 

This daunting multiplicity of forms extends to the atomic-physics problem, as  
well.  The Barker-Glover \cite{22} calculation of $(Z {\alpha})^4$  
corrections incorporated the Darwin-Foldy part of the charge density  
as a recoil correction of order $\left( 1/M^2 \right)$. This is most easily
seen by examining the expression that serves as the baseline for defining
the Lamb-shift energy\cite{22x}. Writing
$$
f(n,j) \equiv \left( 1 + {{(Z\alpha)^2}\over{\left[n - j -\frac{1}{2}
+ \sqrt{(j + \frac{1}{2})^2 - (Z\alpha)^2}\right]^2}}\right)^{-1/2} \,
, \eqno (7a) $$
then for the state of an electron of mass $m_e$ specified by
quantum numbers $(n,l,j)$,
we have to order $(Z \alpha)^4 / M^2$ for the two-body Coulomb problem
$$
E_{nlj}=m_e + M + \mu [f(n,j)-1] -\frac{\mu^2}{2(m_e + M)}[f(n,j)-1]^2
+
\frac{(Z \alpha)^4 \mu^3}{2 n^3 M^2}(\frac{1}{j+\frac{1}{2}} -\frac{1}
{l+\frac{1}{2}})[1-\delta_{l0}], \eqno (7b)
$$
where $\mu$ is the usual reduced mass. This equation can be rewritten as
$$
E_{nlj}=m_e + M + \mu [f(n,j)-1] -\frac{\mu^2}{2(m_e +
M)}[f(n,j)-1]^2 +
\frac{(Z \alpha)^4 \mu^3}{2 n^3 M^2}(\frac{1}{j+\frac{1}{2}} -\frac{1}
{l+\frac{1}{2}}) + E_{\rm DF}\, , \eqno (7c)
$$
where the contribution of the proton Darwin-Foldy ($\delta_{l 0}$) term to 
the atom's energy is
$$
E_{\rm DF}= \frac{(Z \alpha)^4 \mu^3}{2 n^3 M^2}\delta_{l0}\, . \eqno (7d)
$$
The standard expression\cite{1} for the leading-order nuclear-finite-size
correction to the atom's energy is
$$
E_{\rm FS}= \frac{2 (Z \alpha)^4 \mu^3}{3 n^3}\langle r^2 \rangle_{\rm ch}
\, \delta_{l0}\, , \eqno (7e)
$$
and using Eq.\ (5c) for $\langle r^2 \rangle_{\rm ch}$ in Eq.\ (7e) precisely
reproduces Eq.\ (7d). Consequently, the DF term in an atom can 
be alternatively considered as part
of a recoil correction of ${\cal O}(1/M^2)$ (Eq.\ (7b)) or as the energy shift
due to {\bf a part} of the mean-square radius of the nuclear charge
distribution (Eq.\ (7e)).

Thus, this same Darwin-Foldy term
is {\bf by convention} a recoil correction in atomic physics (viz., 
the Barker-Glover formula, Eq.\ (7b)) and a kinematical factor in electron 
scattering (viz., the Rosenbluth formula, Eq.\ (3)).  This is perfectly  
allowable but somewhat confusing, since that term is part of the charge 
density of the proton in both cases.  It is unfortunately far too late to 
change these conventions for the hydrogen atom.  We do not recommend, however, 
that they be extended to other nuclei. These options were extensively  
discussed many years ago in the nuclear context \cite{4} and are  
clearly formalism dependent (i.e., a theorist's choice).

Equation (7b) was originally developed for the proton, but has been applied to 
other nuclei. For the deuteron problem
Pachucki and Karshenboim\cite{5} have argued that the DF term for a pointlike
deuteron vanishes, and hence $E_{\rm DF}$ should be dropped 
from Eq.\ (7c). Khriplovich, Milstein, and Sen'kov\cite{6} responded that 
only the fortuitous choice in Ref. \cite{5} of a particular g-factor for
the deuteron caused that term to vanish, and in general such a term exists. We 
agree with Ref.\cite{5} that this DF term should not be included in Eq.\ (7c), 
but for different reasons. As we argue below (and as noted in Ref. \cite{6}), 
the choice of inclusion or not is formalism dependent, although in general the 
term is not vanishing. Any such term is a part of the nuclear charge density 
(see the discussion below Refs.\ \cite{8,22}), and contributes a part of the 
mean-square radius of that density. Indeed, as we have seen, whether the 
proton's DF term is a recoil correction or a nuclear-finite-size shift is also 
formalism dependent, although its inclusion in the standard expression (7b) is 
sanctioned by decades of consensus. We strongly advocate that nuclear DF terms 
be included as part of $\langle r^2 \rangle_{\rm ch}$.

We examine electron scattering from the deuteron, \H3, \He3, and \4He in turn 
using Eq.\ (3) \cite{7}. This is particularly relevant and topical because of 
the recent re-analysis of the experimental electron-deuteron scattering data 
by Sick and Trautmann \cite{9}. Their derived radius, $\langle r^2 \rangle_{\rm 
ch}^{1/2} = 2.128(11)$ fm, is the rms radius of the complete deuteron charge 
density. This is typical of most nuclear calculations, which work with the 
charge density using the invariant convention (although there are some  
exceptions). 

The deuteron has Z=1 and spin 1, which adds another form factor to the 
``charge-like'' form factor, $G_1$, and ``magnetic-like'' form factor, $G_2$:  
the ``quadrupole-like'' form factor, $G_3$.  Various definitions and 
combinations can be used, and we use the notation and definitions of 
Refs.\ \cite{24,25}. Because the charge-monopole (the spherical part of $\rho$) 
and charge quadrupole (the nonspherical part of $\rho$) contributions are  
incoherent (unless the deuteron spin is somehow constrained), 
the $A_0$ function of Eq.\ (3) becomes
$$
A_0 (Q^2) = G^2_C + \frac{8}{9} \left[ \frac{Q^{2} G_{Q}}{4  
M^{2}} \right]^2, \nonumber \eqno (8a)
$$
where for small $Q^2$ the charge form factor $G_C$ is approximately\cite{25}
$$
G_C (Q^2) \simeq G_1 + \frac{Q^{2}}{6} Q_d\, , \eqno (8b)
$$
while the quadrupole form factor $G_Q$ depends on $G_1, G_2$, and 
$G_3$\cite{25}. The static deuteron quadrupole moment is $Q_d = 0.286\, 
{\rm fm}^2$. Equation (8b)  
is equivalent to corresponding forms in Refs. \cite{5,6,25,26,27}. Defining
$\langle r^2 \rangle_{\rm ch} \, = \, -6\, G_C^{\prime} (0)$ and $\langle r^2 
\rangle_1 = -6\, G^{\prime}_1 (0)$, one finds
$$
\langle r^2 \rangle_{\rm ch} =  \langle r^2 \rangle_1 - Q_d\, . \eqno (8c)
$$
Note that $\langle r^2 \rangle_{\rm ch}$ is the mean-square charge radius, and
not $\langle r^2 \rangle_1$; $-Q_d$ provides a  
Darwin-Foldy-type correction to $G_1$, and is only one part of $\langle  
r^2 \rangle_{\rm ch}$. Because there are alternative form factor definitions 
for the deuteron, there are corresponding alternative size definitions. However,
$\langle r^2 \rangle_{\rm ch}$ is both unique and physically motivated.

The \H3 and \He3 cases (both having spin - $\haf$) mirror
the treatment of the proton, as in Ref.\ \cite{10}, where their $F_C (Q^2)$ is  
the analogue of $G_E$ in Eq.\ (2) and $F_C/\eta^{1/2}$ is the complete  
charge form factor in the invariant representation.  Reference \cite{11}, on  
the other hand, uses a charge operator normalized according to the
covariant convention and their form factor denoted $F_{\rm ch} (Q^2)$ 
differs from that of Ref.\ \cite{10} by an additional factor of 
$\eta^{1/2}$ [$F_{\rm ch}/\eta$ is the charge form factor if one uses the 
invariant normalization convention].  The mean-square charge radius obtained
from Ref.\ \cite{10} is therefore given by ($-6\, F^{\prime}_{\rm C} (0) + 
3/4 M^2)$, while from Ref.\ \cite{11} it is ($-6\, F^{\prime}_{\rm ch} (0) + 
3/2 M^2)$.

For completeness we also consider the spinless nucleus, \4He . The form factor 
and the invariant form of the charge operator for a spin-0 nucleus are the same 
to order $(v/c)^{2}$, and there are no DF corrections. We find\cite{5,6,7,8,16}
$B_0 = 0$,

$$
\rho =  \frac{(E^{\prime} + E)}{\sqrt{4E^{\prime}E}} F_0 (Q^2) 
\simeq F_0  (Q^2) \left( 1 + {\cal{O}} \left( \frac{1}{M^4} \right) 
\right) \, , \eqno (9)
$$
and $\langle r^2 \rangle_{\rm ch} = -6\, F_0^{\prime} (0)$, which is another 
attractive property of the invariant form.

Manifest covariance, which emphasizes form factors, is the traditional way to
implement special relativity, but it is not the only one.
Lorentz invariance (at least to  
order $(v/c)^2$, which is the limit of our interest here) can be implemented  
by constructing explicit many-body representations of the Poincar\'e  
group \cite{8,12,27}.  In this scheme, no part of the charge density is  
more fundamental than any other.  Rather, one works with the  
complete density, including ``boost'' effects such as the Thomas  
precession\cite{28,8}.  For these reasons (based on common  
nuclear practice) we strongly recommend the convention that the mean-square 
radius of the complete nuclear charge distribution be used when computing 
energy shifts. This further implies that no ``Darwin-Foldy'' pieces of the 
mean-square charge radius of a nucleus should be incorporated into ``recoil'' 
corrections.  If the latter is nevertheless done, it is imperative  
that this convention be stated explicitly.

\begin{figure}[htb]
\epsfig{file=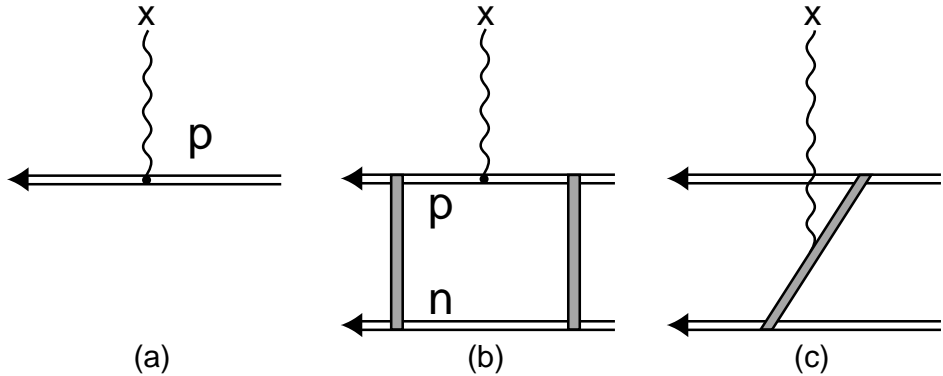,height=2.00in,bbllx=70pt,bblly=550pt,bburx=475pt,
bbury=700pt}
\caption{Deuteron and proton interactions with external electric field
(curly line). The nucleons are depicted as double lines, while meson exchanges
in deuterium that lead to binding or electric currents are shown as shaded 
double lines connecting the proton and neutron. Figure (1a) shows the proton, 
(1b) shows the deuteron graph that generates the ``matter'' radius, while (1c)
illustrates meson-exchange currents. The graph showing the neutron's 
finite-size contribution (identical to Fig.\ (1b) with the curly line attached 
to the neutron) is not shown.}
\end{figure}

Whatever conventions are adopted for the proton, consistency within the 
framework of nuclear physics (which treats nuclei as composed of nucleons)
requires that the physics of the deuteron (or any heavier nucleus) 
incorporate Eq.\ (6). There will be other mechanisms allowed by the presence 
of additional nucleons, as well.
Figure (1a) shows schematically the interaction of a single proton  
with an external Coulomb field.  The solid dot on the double line (the  
proton) indicates the proton's (finite) charge density.  An identical  
interaction occurs in Fig.\ (1b) on that proton inside the deuteron, 
where again the solid dot indicates the full proton charge distribution 
including the DF term. We have indicated by shaded vertical bars on left 
and right the strong interactions that bind the proton and neutron together 
to make a deuteron.  In addition to the proton interaction, the neutron 
has a finite size that contributes via Eq.\ (5b) [note that $\langle r^2 
\rangle_{\rm DF}$ vanishes for a system with no net charge]. The external field
can attach to the neutron in Fig.\ (1b) in an identical fashion to the
proton interaction.
As well,
the spin-orbit-interaction \cite{7,8} (last) term in Eq.\ (6) generates a 
small relativistic correction $\langle r^2 \rangle_{\rm SO}$ in the bound 
deuteron (or any complex nucleus). Figure (1c) illustrates a generic
contribution of the meson-exchange-current (denoted by MEC) 
type\cite{23}, where the flow of mesons that binds the deuteron generates a 
small contribution of relativistic order to the nuclear charge density\cite{12}.

Putting everything together, we can write for the deuteron
$$
\langle r^2 \rangle_{\rm ch} = \langle r^2 \rangle_m +  
\langle r^2 \rangle^n_{\rm ch} + \langle r^2 \rangle^p_{\rm ch} + \langle r^2  
\rangle_B\, , \eqno (10a)
$$
or, equivalently,
$$
\langle r^2 \rangle_{\rm ch} = \langle r^2 \rangle_{\rm pt} +
\langle r^2 \rangle^n_{\rm ch} + \langle r^2 \rangle^p_{\rm ch} \, , \eqno (10b)
$$
where the part due to the binding mechanism is given by
$$
\langle r^2 \rangle_B = \langle r^2 \rangle_{\rm SO} + 
\langle r^2 \rangle_{\rm MEC} + \cdots\, , \eqno (10c)
$$
and the ``point-nucleon'' radius of the deuteron is defined to be
$$
\langle r^2 \rangle_{\rm pt} = \langle r^2 \rangle_m
+ \langle r^2 \rangle_B\, . \eqno (10d)
$$
The nucleon mean-square charge radii are given by Eq.\ (5b) [recall that 
$\langle r^2 \rangle_{\rm DF} = 0$ for the neutron case]. In addition,
$\langle r^2 \rangle_m$ is the mean-square ``matter'' radius, obtained  
directly from the square of the deuteron wave function 
[$\langle r^2 \rangle_m \equiv \int d^3 r | \Psi_d (r)|^2 \, (r/2)^2$,
where $r/2$ is the distance from the deuteron center-of-mass to the proton]. 
Equation (10) is quite general and applies to an arbitrary nucleus
if a factor of $N$ (the number of neutrons) multiplies $\langle r^2  
\rangle^{n}_{\rm ch}$ and a factor of $Z$ (the number of protons)  
multiplies $\langle r^2 \rangle_{\rm ch}$, $\langle r^2 \rangle_{\rm m}$,
and $\langle r^2 \rangle^p_{\rm ch}$.  The correction due to nuclear  
binding mechanisms, $\langle r^2 \rangle_B$, has been written as the sum of  
spin-orbit contributions from the individual neutrons and protons via  
the last term in Eq.\ (6) and (potential-dependent) meson-exchange  
currents, plus \ldots\, . Its presence makes Eqs.\ (10) a definition.

In the traditional interpretation of the isotope shift\cite{1}, one calculates
$\langle r^2 \rangle_{\rm ch} - \langle r^2 \rangle^p_{\rm E}$ as the measure 
of the finite-size difference in the isotope shift, where the first (deuteron) 
term incorporates a proton DF term while the second (proton) term does not.
This difference then includes a term $\langle r^2 \rangle_{\rm DF}$ 
from the proton in the deuteron that counterbalances a similar term implicit
in the Barker-Glover recoil correction for the proton contained in Eq.\ (7b). 
This has been done consistently\cite{1}. Thus, the proton-size 
effect (including the DF part) completely cancels in the $d - p$ isotope shift. 
This cancellation must occur on physical grounds (see Fig.\ (1)), irrespective 
of the fact that in the proton case {\bf by convention} we choose to call the 
DF term a ``recoil'' correction, rather than a finite-size term. 

At the level of accuracy of Ref.\ \cite{3}, however, this approach is no longer 
adequate. Each nuclear finite-size effect comes with its own
reduced-mass correction (see Eq.\ (7e)). The proton 
finite-size corrections in the deuterium atom and in the hydrogen atom differ
by ~0.9 kHz in the 2S-1S isotope shift from this effect, although it is very 
tiny for the DF part alone. The finite-size correction should be calculated 
for each isotope with the proper reduced mass before they are subtracted.

\begin{table}[htb]
\centering
\caption{Calculation of the deuteron rms matter radius for a variety of 
potential models listed on the left. The full radius for each potential is 
shown in the first column of numbers, followed by the zero-range 
approximation for that case, and the 
defect mean-square radius (the difference in the squares of those 
columns). The final column combines the defect with the ``experimental'' 
value\protect\cite{37} of the zero-range approximation (1.9847(18) fm) to 
obtain a prediction for the full matter radius.}

\hspace{0.25in}

\begin{tabular}{|l|cccc|} \hline

\multicolumn{1}{|c|}{\rule{0in}{3ex} Potential Model}&
\multicolumn{1}{c}{$\langle r^2 \rangle^{1/2}\, ({\rm fm})$}&
\multicolumn{1}{c}{$\langle r^2 \rangle_{\rm ZR}^{1/2}\, ({\rm fm})$}&
\multicolumn{1}{c}{$\Delta \langle r^2 \rangle\, ({\rm fm^2})$}&
\multicolumn{1}{c|}{$\langle r^2 \rangle_{\rm m}^{1/2}\, ({\rm fm})$}\\[0.5ex] 
\hline \hline
\multicolumn{5}{|c|}{Second-Generation Potentials}\\ \hline

Nijmegen (full-rel)& \rule{0in}{2.5ex}1.9632  &  1.9811  &  -.0705  & 1.9669 \\
Nijmegen (nl-nr)      &  1.9659  &  1.9831  &  -.0681  &  1.9675 \\
Nijmegen (nl-rel)     &  1.9666  &  1.9839  &  -.0683  &  1.9675 \\
Nijmegen (loc-nr)     &  1.9671  &  1.9843  &  -.0680  &  1.9675 \\
Nijmegen (loc-rel)    &  1.9675  &  1.9847  &  -.0680  &  1.9675 \\
Reid Soft Core (93)   &  1.9686  &  1.9866  &  -.0709  &  1.9668 \\
Argonne V$_{18} $     &  1.9692  &  1.9865  &  -.0685  &  1.9674 \\ \hline  

\multicolumn{5}{|c|}{First-Generation Potentials} \\ \hline

Reid Soft Core (68) \rule{0in}{2.5ex}& 1.9569  &  1.9683 & -.0446  & 1.9735 \\ 
Bonn (CS)                & 1.9687  &  1.9871 &  -.0726  &  1.9664 \\
Paris                    & 1.9714  &  1.9890 &  -.0695  &  1.9672 \\
de Tourreil-Rouben-Sprung& 1.9751  &  1.9926 &  -.0694  &  1.9672 \\
Argonne V$_{14}$         & 1.9816  &  2.0005 &  -.0754  &  1.9657 \\
Nijmegen (78)            & 1.9874  &  2.0069 &  -.0780  &  1.9650 \\
Super Soft Core (C)      & 1.9915  &  2.0119 &  -.0816  &  1.9641 \\ \hline 

\end{tabular}
\end{table}

Our final topic is a preliminary analysis of the deuteron charge radius  
in the non-relativistic impulse approximation\cite{23} (i.e., the ``matter''  
radius).  The zero-range approximation\cite{29} results from neglecting the 
d-state wave function and replacing the deuteron reduced s-state wave function 
by its asymptotic form, $A_S e^{- \beta r}$, where $\beta$ is the deuteron 
relativistic wave number and $A_S$ is the s-wave asymptotic normalization 
constant. This excellent approximation overestimates $\langle r^2 
\rangle^{1/2}$ by less than 1\%. Table 1 shows a calculation of  
$\langle r^2 \rangle^{1/2}$ for a wide variety of first-generation  
\cite{30,31,32,33,34,35,36} (i.e., older) and second-generation potentials
\cite{13,14,15} (i.e., newer ones that fit the nucleon-nucleon scattering  
data from very well to exceptionally well).  The full $\langle r^2  
\rangle^{1/2}$ is followed by the zero-range result for that potential.   
The residual, $\Delta \langle r^2 \rangle = \langle r^2 \rangle - \langle  
r^2 \rangle_{\rm ZR}$, is next.  The residual is small and for our  
second-generation potentials spans the range: $-0.0695(15) \; {\rm fm}^2$.  
The zero-range result using the best current values of $A_S$  
$(0.8845(8)\; {\rm fm}^{-1/2})$ and $\beta$ \cite{37} is 
$\langle r^2 \rangle_{\rm ZR} = A^2_S/(16  \beta^3) = (1.9847(18)
\; {\rm fm})^2 $, which combines with the residual just quoted to give our best 
theoretical value for the root-mean-square (rms) matter radius of the deuteron:
$$
_{\rm th}\langle r^2 \rangle^{1/2}_m = 1.967(2)\; {\rm fm}\, . \eqno (11)
$$

\begin{table}[htb]
\centering
\caption{Experimental and theoretical 2S-1S deuterium-hydrogen 
isotope shifts in kHz. The experimental value is given on the left, 
followed by the theoretical value for point nuclei
(with no Darwin-Foldy terms included in either non-radiative recoil 
contribution), the sum of nuclear polarization, nuclear Lamb shift 
and higher-order Coulomb finite-size contributions is next, followed 
on the right by the leading-order nuclear finite-size contribution
(including all nuclear Darwin-Foldy terms) adjusted to produce agreement with
the experimental isotope shift.}

\hspace{0.25in}

\begin{tabular}{|c||ccc|}
\hline
experimental & point nuclei & misc. nuclear& nuclear size \\ \hline
670 994 334(2) \rule{0in}{2.5ex}& 670 999 503.2 & 19.2 & -5188.4  \\ \hline

\end{tabular}
\end{table}

This result is our baseline, from which deviations signal ``exotic''
components of the deuteron charge density.  We can make our own estimate of 
this deviation by using the current experimental value\cite{3}
of the 1S-2S isotope shift: 670 994 334(2) kHz.  We also use an  
updated version of the theoretical analysis presented in Ref.\ \cite{1},
which is displayed in Table 2.  We use the improved $\frac{m_p}{m_e}$ ratio  
of Ref.\ \cite{38} (1836.1526665(40)) and the  
$\frac{m_d}{m_p}$ ratio of Ref.\ \cite{39} (1.9990075009(8)).   
We also use the improved deuteron polarizability of Ref. \cite{40};
the proton polarizability of Ref.\ \cite{41} cancels in the isotopic 
difference.  Higher-order $(Z \alpha)^5$ and $(Z \alpha)^6$ Coulomb 
finite-size corrections are obtained from Ref.\ \cite{42}. The neutron 
mean-square charge radius is taken from Ref.\ \cite{43}: 
$\minus 0.1140(26)~{\rm fm}^2$.
All other constants are taken from Ref.\ \cite{44}. Using the deuteron 
mean-square charge radii defined by Eq.\ (10), we obtain the experimental value
of the deuteron point-nucleon radius
$$
_{\rm exp}\langle r^2 \rangle_{\rm pt}^{1/2} = 1.9753(11)\; {\rm fm}\, , 
\eqno (12)
$$
and
$$
_{\rm exp}\langle r^2 \rangle_{\rm pt}^{1/2} -\, {_{\rm th}\langle r^2 
\rangle_{\rm m}^{1/2}} = 0.008(2)\; {\rm fm}\, , \eqno (13)
$$
where the error in Eq.\ (12) is obtained by compounding a 1.5 kHz 
$\frac{m_p}{m_e}$ uncertainty, the 2 kHz experimental uncertainty, an 
estimated 4 kHz uncertainty in QED calculations\cite{1}, and an (equivalent) 
3.5 kHz uncertainty from 
the neutron charge radius. These results are shown in Table 3. On the scale of 
these uncertainties the DF terms discussed earlier are very large for the 2S-1S
transition, approximately 45 kHz/$A^2$ ($A$ is the nucleon number), where 
roughly 5 kHz changes $\langle r^2 \rangle^{1/2}_{\rm ch}$ by 0.001 fm.

\begin{table}[htb]
\centering
\caption{Experimental and theoretical deuteron radii. The deuteron matter radius
corresponding to second-generation nuclear potentials renormalized to the
experimental zero-range approximation and the experimental point-nucleon 
charge radius of the deuteron are 
shown in the first two columns, followed by the difference of experimental 
and theoretical results. Relativistic corrections to the mean-square 
charge radius from the electromagnetic spin-orbit interaction and from MEC 
(assuming minimal nonlocality) are listed in the next two columns. The final
theoretical estimate of the charge radius for pointlike nucleons is listed in 
the sixth column. No uncertainty is given in the final estimate because of 
consistency problems between the MEC and the nuclear potentials.}

\hspace{0.25in}

\begin{tabular}{|cc|c||cc|c|}
\hline
$_{\rm th}\langle r^2 \rangle_{\rm m}^{1/2}\, ({\rm fm})$ &
$_{\rm exp}\langle r^2 \rangle_{\rm pt}^{1/2}\, ({\rm fm})$ &
Diff.\ (fm) &
$\langle r^2 \rangle_{\rm SO}\, ({\rm fm^2})$ &
$\langle r^2 \rangle_{\rm MEC}\, ({\rm fm^2})$ &
$_{\rm th}\langle r^2 \rangle_{\rm pt}^{1/2}\, ({\rm fm})$ \\ \hline
1.967(2) \rule{0in}{2.5ex}& 1.9753(11) & 0.008(2) 
& -0.0014 & 0.0159 & 1.971\\ \hline

\end{tabular}
\end{table}

The atomic results above can be contrasted with the less precise determination 
of $\langle r^2 \rangle^{1/2}_{\rm pt}$ using Eqs.\ (10) and the 
electron scattering results of Refs. \cite{9,20x}:
$$
_{\rm exp}\langle r^2 \rangle^{1/2}_{\rm pt} = 1.966(13)\; {\rm fm}\, , 
\eqno (14)
$$
from which we obtain
$$
_{\rm exp}\langle r^2 \rangle_{\rm pt}^{1/2} -\, {_{\rm th}\langle r^2 
\rangle_{\rm m}^{1/2}} = -0.001(13)\; {\rm fm}\, . \eqno (15)
$$
At this level of precision, the result (15) is null. Equations (10b) and (12)
lead to a full deuteron charge radius from the isotope shift of 2.136(5) fm,
which is consistent with the value of 2.128(11) fm from Ref. \cite{9}.

Although the result (13) is effectively nonzero, there is one caveat about its 
significance.  The matter radius derived earlier is not entirely well  
defined.  It was shown long ago \cite{12} that to order $(v/c)^2$ there
are 2 unitary equivalences that arise naturally in treating relativistic  
corrections; these are the (pion) chiral-rotation equivalence  
specified by a parameter, $\mu$, and the quasi-potential equivalence  
(similar to electromagnetic gauge-dependence) specified by a  
parameter, $\nu$.  These parameters modify the nuclear potential  
through nonlocal terms, and also modify the nuclear charge operator  
through meson-exchange currents.  Because none of the representations 
corresponds precisely to a nonrelativistic (i.e., momentum-independent)
potential, no  
specification of $\mu$ and $\nu$ is possible without performing a  
consistent relativistic calculation (at least to order $(v/c)^2$).   
Since a unitary transformation cannot change observables (and  
hence the zero-range approximation is unchanged), only the defect  
wave function and the defect mean-square radius $(\langle r^2 \rangle_m -  
\langle r^2 \rangle_{\rm ZR})$ can be changed and both will therefore  
depend on $\mu$ and $\nu$, as will $\langle r^2 \rangle_{\rm MEC}$.   
Both ($\langle r^2 \rangle_{\rm m} + \langle r^2 \rangle_{\rm MEC}$) and 
$\langle r^2 \rangle_{\rm ch}$ do not.  We can stipulate  
conditions on the potential that will restrict the parameters $\mu$ and  
$\nu$.  One condition is ``minimal nonlocality'', which requires the  
nuclear tensor force to be as local as possible and the entire force to be  
energy independent.  This is equivalent to $\mu = 0$ and  
$\nu = 1/2$\cite{12}, and bears a rough correspondence to Coulomb gauge in  
atomic physics.  Such a representation is probably the closest to (but  
not quite the same as) using the local potentials that are the norm in  
nuclear physics.  This representation for the MEC charge operator is  
well known \cite{12} and produces
$$
\langle r^2 \rangle_{\rm MEC} 
\left|_{\stackrel{\hspace*{-0.15in} \mu = 0}{\nu= 1/2}} 
\simeq 0.0159\, {\rm fm}^2 \, , \right. \eqno (16)
$$
and together with
$$
\langle r^2 \rangle_{\rm SO} \simeq - 0.0014\, {\rm fm}^2 \, , \eqno (17)
$$
one finds the full radius
$$
_{\rm th}\langle r^2 \rangle^{1/2}_{\rm pt} = 1.971\, {\rm fm} \, , \eqno (18)
$$
which makes up approximately half the difference between the experimental  
value and the baseline estimate predicated on nonrelativistic second-generation 
potentials: $(\langle r^2 \rangle^{1/2}_{\rm pt} - \langle r^2 \rangle^{1/2}_m)$
given in Table 3.  Hopefully, the remaining .004 fm comes from the difference 
between a true relativistic treatment of the deuteron and our nonrelativistic 
one that we have supplemented with (somewhat) {\it ad hoc} corrections. Our 
results for $\langle r^2 \rangle_{\rm B}$ are similar to those of Ref.\ 
\cite{46}.

In summary, we have reviewed the various ways that nuclear sizes are  
incorporated into electron scattering and atomic calculations.  We strongly 
recommend the convention that {\bf complete} nuclear charge radii be  
used in calculating atomic energy shifts, rather than radii based on  
arbitrary form factor definitions.  A ``baseline'' value of the deuteron rms 
radius was calculated using nonrelativistic second-generation potentials to 
correct the (excellent) zero-range approximation.  A value of the deuteron rms  
radius extracted from the d-p isotope shift is .008(2) fm  
larger than this baseline value, some of which is almost certainly due to  
meson-exchange currents.  A complete resolution of the problem caused by this 
difference awaits a relativistic treatment of the deuteron dynamics\cite{47} 
that is of ``second-generation'' quality, because we are dealing 
with very small size differences.

\noindent {\bf Acknowledgements}

The work of J.\ L.\ F.\ was performed under the auspices of the United  
States Department of Energy. D.\ W.\ L.\ S.\ is grateful to NSERC Canada
for continued support under Research Grant No.\ SAP00-3198. The work of 
J.\ M.\ is supported under Grant No.\ PB94-0900 of DGES, Spain.
We would like to thank T.\ W.\ H\"ansch for providing information about
his experiments, I.\ Khriplovich for a useful discussion about g-factors,
and K.\ Pachucki for useful comments on an early version of this manuscript.


\begin{thebibliography}{99}
\bibitem{41} I.\ B.\ Khriplovich, and R.\ A.\ Sen'kov, nucl-th/9704043. Just
as the proton size cancels in the d-p isotope shift, the proton polarizability
also occurs in the deuteron, and hence will cancel in the isotope shift. The 
neutron polarizability in the deuteron has not been included in our 
calculations.
\end{thebibliography}
\end{document}